\documentclass[proof]{WileyASNA-v1}
\usepackage{comment}
\usepackage{pdfpages}

\articletype{Article}%

\received{XXX}
\revised{YYYY}
\accepted{ZZZ}

\begin{document}

\title{Seeing-limited Coupling of Starlight into Single-mode Fiber with a Small Telescope }

\author[1]{David H. Sliski}
\author[1]{Cullen H. Blake}
\author[2]{Jason D. Eastman}
\author[3]{Samuel Halverson}

\authormark{Sliski \textsc{et al.}}

\address[1]{\orgdiv{Department of Physics and Astronomy}, \orgname{University of Pennsylvania}, \orgaddress{\state{Pennsylvania}, \country{USA}}}

\address[2]{\orgdiv{Center for Astrophysics \textbar \ Harvard \& Smithsonian}, \orgaddress{\state{Massachusetts}, \country{USA}}}

\address[3]{\orgdiv{Jet Propulsion Laboratory \textbar \ California Institute of Technology}, \orgaddress{\state{California}, \country{USA}}}

\corres{Cullen H. Blake \\
\email{chblake@sas.upenn.edu}}

\presentaddress{Department of Physics and Astronomy, University of Pennsylvania, Philadelphia, PA 19103}

\abstract{An optical fiber link to a telescope provides many advantages for spectrometers designed to detect and characterize extrasolar planets through precise radial velocity (PRV) measurements. In the seeing-limited regime, a multi-mode fiber is typically used so that a significant amount of starlight may be captured. In the near-diffraction-limited case, either with an adaptive optics system or with a small telescope at an excellent site, efficiently coupling starlight into a much smaller, single-mode fiber may be possible. In general, a spectrometer designed for single-mode fiber input will be substantially less costly than one designed for multi-mode fiber input. We describe the results of tests coupling starlight from a 70 cm telescope at Mt. Hopkins, Arizona into the single-mode fiber of the MINERVA-Red spectrometer at a wavelength of $\sim$850 nm using a low-speed tip/tilt image stabilization system comprising all commercial, off-the-shelf components. We find that approximately 0.5$\%$ of the available starlight is coupled into the single-mode fiber under seeing conditions typical for observatories hosting small telescopes, which is close to the theoretical expectation. We discuss scientific opportunities for small telescopes paired with inexpensive, high-resolution spectrometers, as well as upgrade paths that should significantly increase the coupling efficiency for the MINERVA-Red system. }


\maketitle

\section{Introduction}\label{sec1}
Over the past fifty years, the precision of stellar radial velocity measurements has improved by more than two orders of magnitude (\citealt{Fischer2016}). Today, a number of precise radial velocity (PRV) instruments recently deployed or being commissioned are capable of $<1$m s$^{-1}$ instrumental precision, and teams around the world are working to mitigate technical challenges on the way toward 10 cm s$^{-1}$ instrumental precision (see, for example, \citealt{kpf2018}, \citealt{Robertson2019}, \citealt{Blackman2020}, \citealt{espresso2020}, \citealt{pepe2021}, \citealt{sagi2018}). Fundamentally, these instruments seek to measure Doppler shifts in stellar spectral features at the sub-femtometer level, an extremely challenging technological feat that requires high spectral resolution (R=$\lambda/\Delta\lambda$>10$^{5}$) and exquisite control of systematic sources of measurement error. Most PRV spectrometers are similar in design, employing an echelle grating as the primary dispersing element, a cross disperser to place a large number  of diffraction orders across a CCD or NIR detector, sub-milliKelvin thermal control of the entire instrument, and a multi-mode fiber from the telescope focal plane to the instrument. The physical size of the focused star (1'' corresponds to 60$\mu$m on the focal plane of a 3~m telescope operating at f/4) and a desire to capture as much starlight as possible sets the size of the multi-mode fiber, which is typically tens of microns in diameter. At fixed spectral resolution and without pupil slicing, the fiber diameter determines the physical sizes of the optical components within the instrument, such as the grating and collimating optics (see, for example, \citealt{bingham1979}). The sizes of these components are a primary driver of total instrument cost, and anecdotal evidence suggests that the current generation of PRV instruments designed for sub-m s$^{-1}$ instrumental precision cost in excess of $\$$10 million US dollars to design and build.

A primary advantage of a fiber link to the spectrometer is that the instrument can be placed in a temperature-controlled environment far from the telescope, which is key to achieving sub-milliKelvin temperature stability (see, for example, \citealt{Robertson2019}).  Changes in the illumination of the fiber at the telescope due to guiding errors translate directly into un-calibratable systematic RV errors. The fiber naturally reduces the impact of these guiding errors compared to a slit-based system, but the scrambling effects of the fiber alone are not sufficient when the goal is high RV precision. Multi-mode fibers also suffer from an effect called modal noise that limits achievable signal-to-noise (\citealt{lemke2011}). This speckle pattern in the output of the fiber results from the phase delays experienced by light of the same wavelength propagating along the fiber in different modes. By combining physical agitation of the fiber with an optical double scrambler to exchange the near and far fields of the fiber, both the illumination and modal noise effects can be reduced by factors of thousands but remain important and place stringent requirements on guiding performance for next-generation PRV spectrometers (\citealt{sirk2018}, \citealt{schwab2018}). Finally, multi-mode fibers suffer from focal ratio degradation, which causes the angular spread of the fiber output to be larger than the acceptance cone of the fiber. This effect requires increasing the diameters of instrument optical components so as not to lose light and is minimized by careful preparation of fiber connections and reducing bends in the fiber. 

Unlike a multi-mode fiber, which can have hundreds of modes at optical wavelengths, a single-mode fiber has only two fundamental, nearly Gaussian modes (one in each polarization - see \citealt{Halverson2015}). This results from the fact that the fiber has a very small core diameter and Mode Field Diameter (MFD), typically a few microns at optical wavelengths for standard glass fibers. This presents a number of important advantages from the point of view of instrument design and cost, as discussed in, for example, \citet{ghasempour2012}, \citet{schwab2012}), \citet{crepp2014}, \citet{nem2016}, \citet{Crass2019}, \citealt{parvi2022}, and \citealt{hispec2022}. The single-mode fiber does not suffer from the modal noise or illumination limitations of the multi-mode fiber and also has no focal ratio degradation. At the same time, the small MFD of the fiber means that high spectral resolution can be achieved with a small collimated beam and near-diffraction-limited spectrograph optics. Depending on the desired wavelength coverage, commercial, off-the-shelf optics in standard diameters (50 or 75~mm) are sufficient for many key spectrometer components, even at a resolution of $R>100,000$. The small input spot size typically means that the spectrograph camera system must magnify, meaning relatively long focal lengths and overall reduced cost and complexity compared to a very fast camera system. 

\begin{figure}[h!]
	\centerline{\includegraphics[width=85mm,height=85mm]{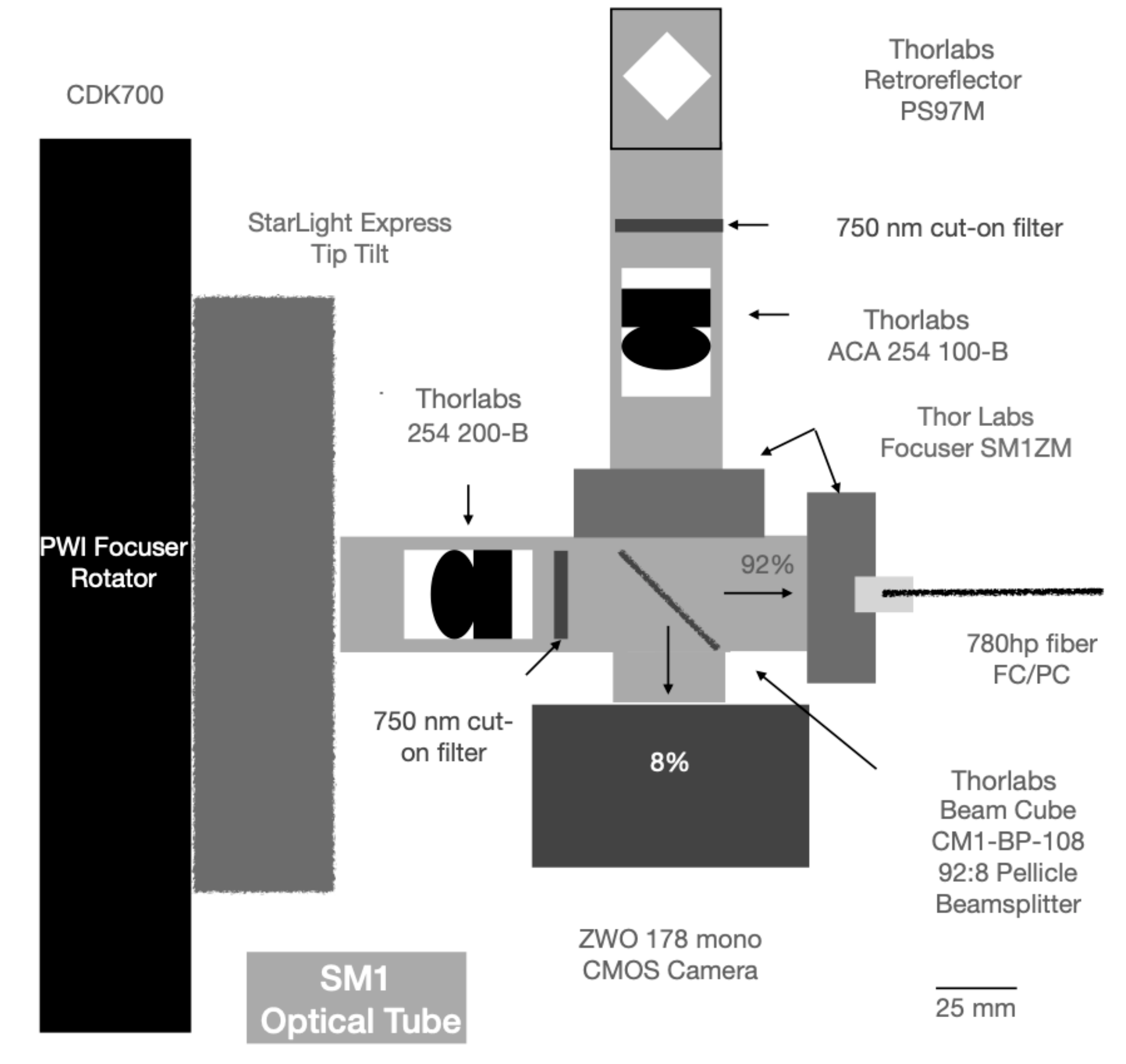}}
	\caption{MINERVA-Red Fiber Acquisition Unit (FAU) with image stabilization guider. The CDK700 telescope is located on the left side, behind the PWI focuser/rotator. The FAU is built from all commercial, off-the-shelf parts and is based on the design described in \citet{bottom2014}.}
	\label{fig:fau}
\end{figure}

The small beam diameter generated by a short focal length collimator and the modest sizes of the individual optical components means that the entire instrument occupies much less volume than an instrument with comparable resolution and multi-mode fiber input, particularly if the traditional instrument does not employ pupil slicing. For example, while the volume of the vacuum chamber housing a traditional multi-mode fiber PRV instrument may be $\sim$~3m$^{2}$, a single-mode instrument with similar resolution and spectral coverage could easily fit within half a cubic meter. Taken together, the smaller component sizes enabled by single-mode fiber dramatically reduce the costs associated with designing and building the instrument. This advantage becomes particularly important when considering high-resolution spectrometers for the next generation of extremely large telescopes, where the seeing-limited instrument may be prohibitively expensive, and the advantages of single-mode fiber instruments operating behind advanced adaptive optics systems are even more significant.

However, coupling starlight into a single-mode fiber is challenging. Given a stellar PSF that is tens of microns across at the telescope focal plane, it is clear that very little light is going to couple into a single-mode fiber with a Gaussian fundamental mode having a MFD of $\sim5\mu$m. Changing the focal ratio of the light to shrink the stellar PSF only helps while the beam falls within the acceptance angle of the fiber, defined by a Numerical Aperture (NA) of $\sim0.1 - 0.15$. The integral of the product of the PSF and the fiber mode gives a rough estimate of the maximum coupling efficiency, which is expected to be low if the fiber MFD is much smaller than the FWHM of the PSF. 

Alternatively, the stellar image can be thought of as a collection of speckles rapidly evolving in time. The number of speckles is approximately $(D/r_0)^2$, where $r_{0}$ is the Fried parameter, which characterizes the atmospheric stability of the site, and $D$ is the telescope diameter. At best, on average, one speckle is going to couple to the single-mode fiber at a time. For the 3~m telescope at f/4 (matched to a fiber with NA=0.12), 1$^{\prime\prime}$ seeing at 500 nm ($r_{0}=0.1$~m), and a fiber MFD of 5$\mu$m, either approximation points to a coupling efficiency $\epsilon\ll1\%$. 

\cite{Shaklan1988} more rigorously consider the expected coupling efficiency into single-mode fibers for the seeing-limited case over a range of $D/r_{0}$. Even assuming high-speed tip/tilt guiding to center the star on the fiber (which would likely require corrections at a rate of tens of Hz to fully compensate for atmospheric motion) and matching the telescope focal ratio to the fiber NA, these authors find that maximum coupled power, including the effects of coupling efficiency and telescope collecting area, peaks at an efficiency of $\epsilon\sim18\%$ at $D/r_{0}\sim4$. This total coupled power falls off for larger telescopes and decreases by a factor of $\sim5$ overall without high-speed tip/tilt guiding. While the MINERVA-Red system does have an image stabilization system, described in Section 2, this system is not capable of the type of high-speed corrections included in the \citet{Shaklan1988} simulations. 

 This suggests two paths forward for employing single-mode fibers to feed light into astronomical spectrometers - feeding the fiber with an Adaptive Optics (AO) system or using a small telescope at an excellent observing site. Given the advantages of single-mode fiber spectrometers, there is growing interest in the community in both approaches. In principle, AO systems can provide a diffraction-limited stellar image that can be coupled with high efficiency into a single-mode fiber. The advantages of AO system and single-mode fibers for high-resolution spectroscopy were recognized in the late 1990s (see, for example, \citealt{jian1998}), and several efforts to build instruments of this type are underway (\citealt{gibson2022, Bechter2016, mawet2019}). Recently, work at Subaru and the Large Binocular Telescope (LBT) has demonstrated coupling efficiency into single-mode fibers up to $50\%$ at near-infrared wavelengths using modern, sophisticated AO systems (\citealt{Bechter2016}, \citealt{nem2016b}, \citealt{nem2017}). In the seeing-limited regime, the RHEA project has demonstrated an inexpensive, high-resolution single-mode fiber spectrometer on a small telescope (\citealt{bento2016}, \citealt{Fager2016}, \citealt{fager2018}). 

\section{MINERVA-Red Spectrometer}\label{sec:mred}

MINERVA-Red is a high-resolution, cross-dispersed echelle spectrometer designed for Doppler measurements of nearby low-mass stars for the detection and characterization of extrasolar planets. The spectrometer covers a spectral region between 820 and 920~nm, where there is relatively little telluric water 
vapor absorption and low-mass stars are relatively bright,  at a resolution of $R\sim45,000$ using a deep depletion CCD detector. The spectrometer is designed around single-mode fiber input (780HP fiber with a MFD at 1/e$^{2}$ of 5$\mu$m, NA=0.13, and a single-mode cutoff wavelength of $\lambda = 780$~nm) and is built primarily from commercial, off-the-shelf components. The optics are housed in a vacuum chamber and thermally controlled enclosure that provides milliKelvin level temperature stability (\citealt{sliski2017}), which enables intrinsic stability of the wavelength solution equivalent to better than 5 m s$^{-1}$. Calibration of residual instrumental drifts is accomplished with observations of a Uranium-Neon lamp through an all-fiber system that enables rapid switching between on-sky observations and observations of calibration sources. The MINERVA-Red spectrometer and on-sky radial velocity measurement performance will be described in detail in a future publication. 

Starlight is coupled into the science fiber with a custom Fiber Acquisition Unit (FAU) attached to a 0.7~m CDK700 telescope built by PlaneWave Instruments. The telescope, which is part of the MINERVA array at the Fred Lawrence Whipple Observatory (FLWO) at Mt. Hopkins, Arizona, has gold-coated mirrors to optimize the throughput in the 820 to 920~nm range. The telescope has a Nasmyth focus that allows the FAU to have a near-constant gravity vector by disabling the telescope's integrated field derotator. The FAU, which is shown in Figure 1, is similar in design to the unit described in \cite{bottom2014}, but with the addition of an image stabilization guider capable of low-speed tip/tilt corrections and a focal reducer. The FAU uses a pellicle beam splitter, which provides very low fringing, to send 8$\%$ of the $\lambda>750$~nm light to a ZWO 178 monochrome CMOS guide camera and 92$\%$ of the light to the science fiber. Backlighting the science fiber, which is attached to the FAU with standard FC/PC connectors, with an 850~nm laser and imaging the spot onto the guide camera through a corner cube and air-spaced doublet lens allows for precise determination of the fiber location. The retro-reflection path also contains a 750~nm long-pass filter to ensure that when that arm of the FAU is manually focused at installation by observing bright stars, the focus is set using light with a wavelength that is close to the wavelength of the science light. The retro-feed process involving back-lighting the science fiber calibrates any movement within the FAU system and is carried out between science exposures. In practice, we find that the measured variations in the fiber location are very small, at the level of a few microns over days. The entire assembly, which is built around SM1 components from Thorlabs, provides intrinsic stability of the fiber location at the $1\mu$m level. 

Optimal coupling of starlight into a single-mode fiber requires matching the stellar PSF to the Gaussian fundamental mode of the fiber and matching the focal ratio of the telescope beam incident on the fiber face to the NA of the fiber. The beam from the CDK700 telescope is nominally f/6.3, and the 780HP fiber has a nominal NA of 0.13 (defined by the 1/e$^{2}$ intensity points), corresponding to f/3.8. The actual NA value may vary between fibers or as a function of wavelength. The optimal f/\# for single-mode fiber coupling will also depend on the details of the stellar PSF shape, which will not be a Gaussian even with a perfect telescope having no central obstruction (in which case the PSF will be an Airy function and optimal coupling to single-mode fiber is always less than $100\%$). A central obstruction reduces the upper limit of coupling efficiency further (see, for example, \citealt{Shaklan1988, nem2017}). 

We aim to increase the efficiency of our seeing-limited coupling by using a Thorlabs air-spaced doublet with a focal length of 200~mm incorporated into the FAU before the beam splitter to approximately match the telescope beam to the fiber NA. This simple optic provides seeing-limited imaging over our narrow wavelength range within a small field of view that is still many times larger than the fiber. With the focal reducer, the star is imaged onto the ZWO 178 guide camera with 2.4$\mu$m pixels at a plate scale of 0.2$^{\prime\prime}$ pixel$^{-1}$ and a focal ratio of f/3.5. At this focal ratio, the single-mode fiber MFD of 5~$\mu$m corresponds to 0.42$^{\prime\prime}$, substantially less than the average natural seeing at virtually all observatories.

\begin{figure}[h!]
	\centerline{\includegraphics[width=85mm,height=60mm]{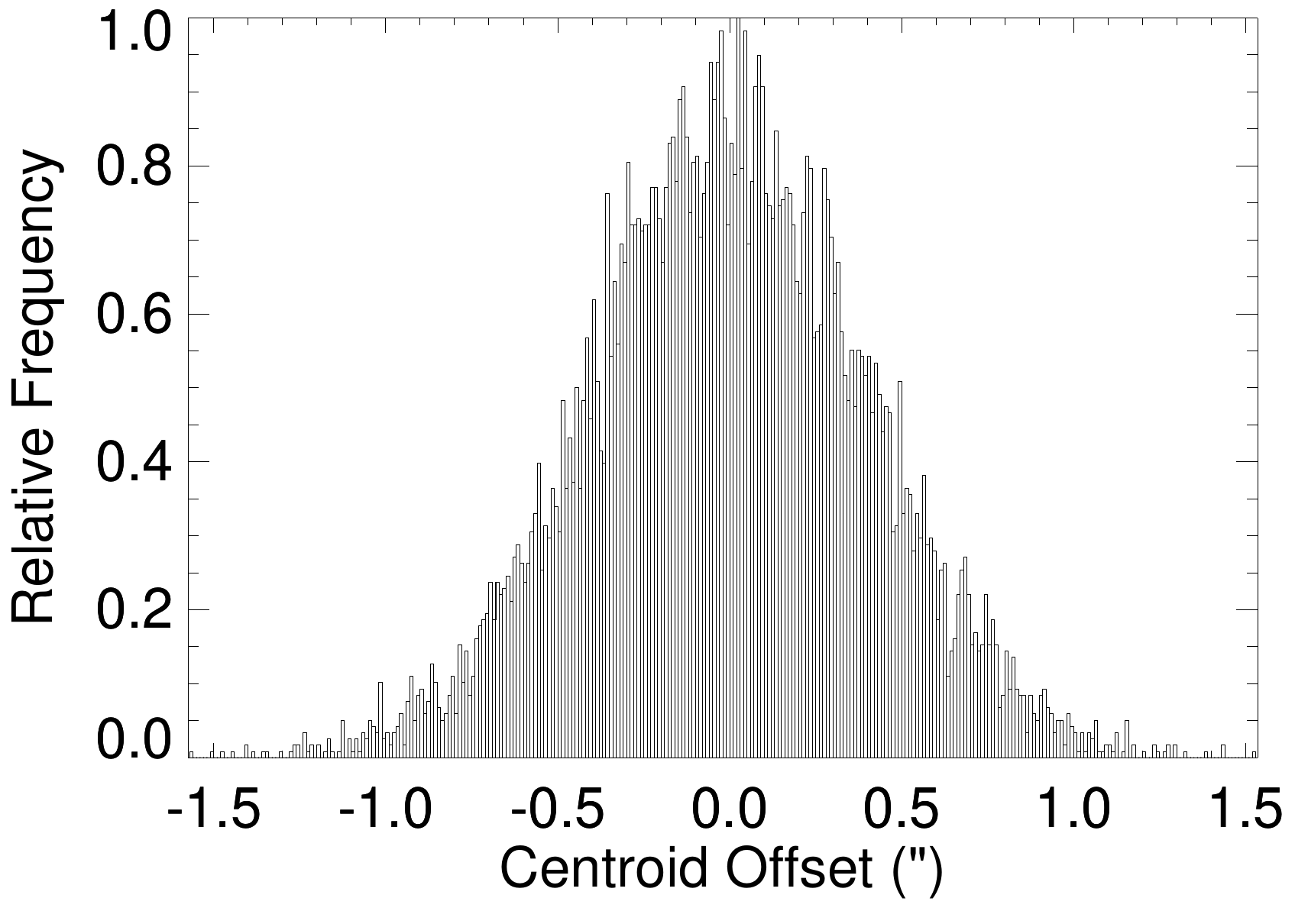}}
	\caption{Measured variation in the position of a stellar centroid on the MINERVA-Red FAU guide camera relative to the average position. With the image stabilization system operating at approximately 0.3 Hz, the image motion is approximately 0.5$^{\prime\prime}$ (rms). The fiber MFD is approximately 0.4$^{\prime\prime}$. Without the image stabilization system, the stellar centroid frequently experiences large departures up to 2$^{\prime\prime}$ or more in windy conditions due to the wind load on the telescope truss.}
	\label{fig:guiding}
\end{figure}

Between the telescope and the focal reducer is an image stabilization guiding module from StarlightXpress. This unit contains a BK7 plate with a VIS AR coating mounted on voice coils that provide motion in two axes to produce parallel displacements of the stellar image. This method of low-speed tip/tilt correction using a plate in transmission is possible because the plate is placed in a relatively slow, converging beam. The total range of motion is $\pm8^{\prime\prime}$ and the unit is intrinsically capable of making corrections at more than 10 Hz. We use a custom software package to drive the image stabilization unit and guide the telescope based on the measured stellar position on the guide camera. If position correction outside the range of the tip/tilt is required, commands are automatically sent to the telescope mount. The transformation between the motion vectors of the tip/tilt and the direction on the sky is measured once by manually commanding tip/tilt moves and fitting the resulting stellar motion to determine the angles that define the orientation of the tip/tilt on the sky. The bandwidth of the guiding loop is currently limited to approximately 1.0 Hz by the time required to process the individual guide camera images, but routinely provides root-mean-square variations (rms)in the stellar centroid $0.5$$^{\prime\prime}$ in typical seeing (1.7$^{\prime\prime}$), which is approximately equal to the angular diameter of the fiber (see Figure \ref{fig:guiding}). The process of target acquisition and initiation of the guide loop is fully automated.

\section{On-sky Coupling Tests}\label{sec:onsky}

The MINERVA-Red calibration system includes an exposure meter that allows for the measurement of the mean light arrival time during extended integrations. This is particularly important for a single-mode fiber spectrometer, where the temporal variations in the coupled starlight are expected to be large (\citealt{Shaklan1988}), translating into uncertainties in the calculation of the barycentric correction (\citealt{eastman2014}). The exposure meter receives 10$\%$ of the science light through an in-fiber 90:10 splitter, and that light is imaged onto a thermoelectrically cooled  Atik 414EX CCD camera using a fiber-coupled doublet lens pair. The exposure meter can be used to estimate the efficiency of coupling starlight into the science fiber by directly comparing contemporaneous measurements of the total stellar flux obtained with the exposure meter and the FAU guide camera. This differential measurement removes uncertainty related to changes in atmospheric transparency and the reflectivity and transmission of the telescope optics from the calculation of coupling efficiency. 

We note that there are optical elements that are not common to both the FAU guide camera and exposure meter paths, and these differences need to be accounted for when estimating the coupling efficiency from the ratio of fluxes on the FAU guide camera and exposure meter. We estimate a small correction factor of order unity using the published transmission curve for the pellicle beam splitter (for un-polarized light) and the published QE curves for the two detectors, integrated from 750~nm to 950~nm. The 780 HP science fiber is AR coated, as are the doublet lens used to image the output of the science fiber onto the exposure meter and the optical windows of the two detectors, so we neglect Fresnel effects from those surfaces. We also neglect losses due to transmission through the fiber, which we expect to be small given the modest fiber length. During testing, we chose to observe hot stars, so the slope of the stellar spectrum is also not included in the calculation of this correction factor.  

We observed the bright star HD 206672 (I $\approx5.0$) over four nights in September and October 2020 for a total of approximately seven hours during MINERVA-Red commissioning. We obtained more than 14,800 2~s images of HD 206672 with the FAU guide camera while measuring the flux on the exposure meter CCD at a rate of approximately 0.75 Hz. These nights were photometric, with variations in the total stellar flux on the FAU camera of $2\%$, but with variable seeing. While the FLWO site has good intrinsic seeing properties and 1.0$^{\prime\prime}$ seeing can be realized at other facilities at the site, during commissioning the mean image quality observed at MINERVA-Red was 1.7$^{\prime\prime}$ with a one standard deviation range of 1.37$^{\prime\prime}$ to 2.03$^{\prime\prime}$ FWHM as estimated by two dimensional Gaussian fits to the stellar images on the FAU guide camera. The delivered image quality is degraded by wind-induced vibrations and loading in the carbon fiber truss of the telescope front end, which is very lightweight, and also by the optical performance and focus of the FAU system itself. In windy conditions, the load on the telescope truss produces substantial image motion but at relatively low frequencies that the active guiding unit can help mitigate. Future upgrades to the MINERVA-Red facility will address these image quality issues, but the current seeing is representative of what is often achievable at sites hosting small telescopes, such as university observatories at low-altitude locations.

\begin{figure}[h!]
	\centerline{\includegraphics[width=85mm,height=60mm]{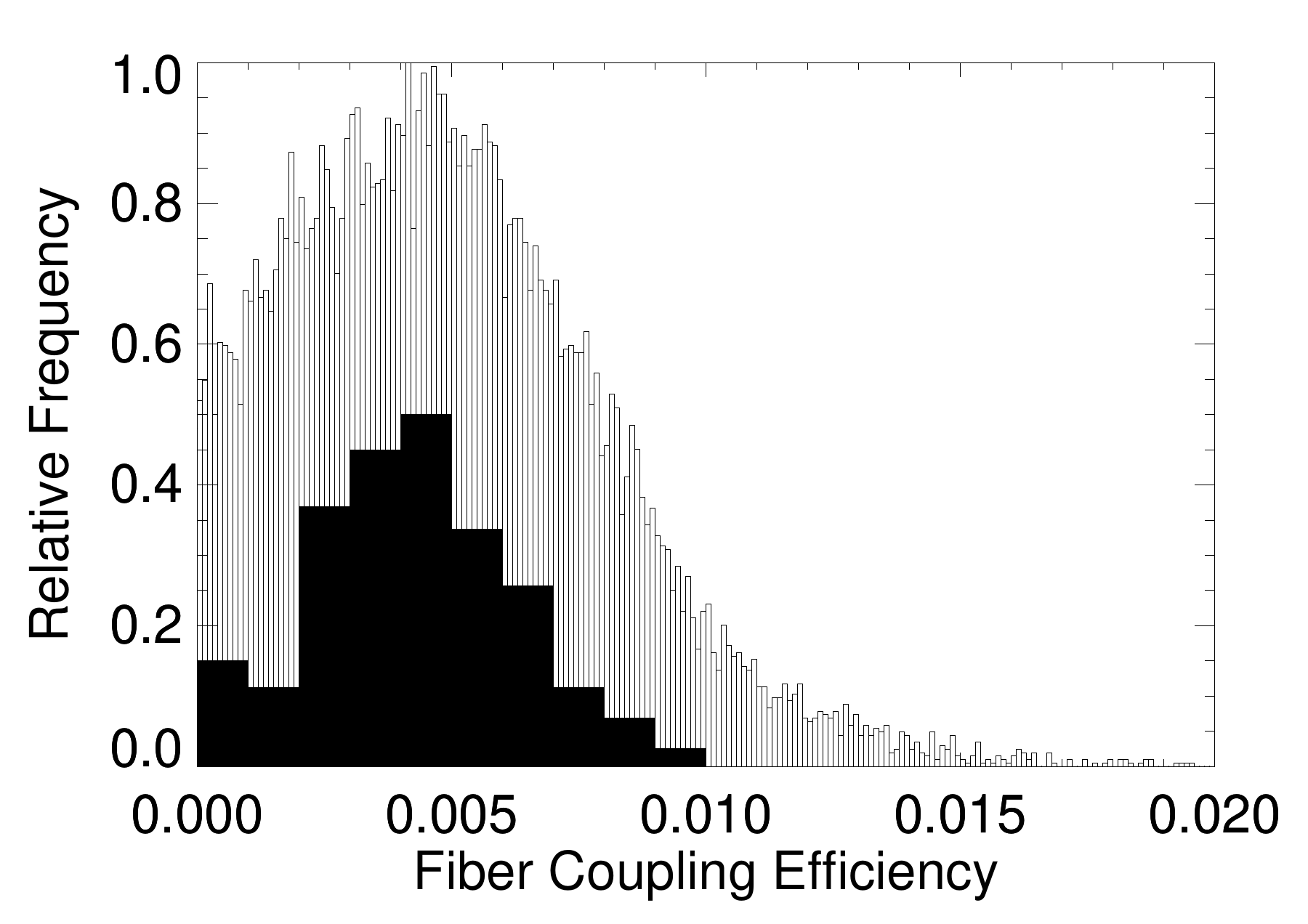}}
	\caption{Measured single-mode fiber coupling efficiency based on 14,890 exposure meter measurements over seven hours of observations under a range of conditions, with a median of  $\epsilon=0.45\%$ (small bins). Solid histogram shows optimal coupling efficiency based on the overlap integral of the stellar PSF as imaged on the FAU guide camera and a Gaussian approximation of the fiber fundamental mode with a MFD of 5 $\mu$m (larger bins, black fill, scaled down for clarity).}
	\label{fig:couple}
\end{figure}

In Figure \ref{fig:couple} we show the overall distribution of the measured coupling efficiency after applying the small correction factor described above, which has a median of $\epsilon=0.45\%$. We compare this to an empirical expectation based on direct integration of the product of the fiber mode and the stellar PSF as imaged on the FAU guide camera, under the assumption that the stellar PSF at the fiber face is the same (which should be the case if the FAU is properly focused). We approximate the fiber mode as a Gaussian with FWHM=MFD/1.68, where MFD is reported by the fiber manufacturer to the 1/e$^{2}$ points. We sum the total flux, after background subtraction, in the product of the fiber mode, centered at the fiber location as determined by the most recent fiber backlighting, and the stellar PSF as imaged on the FAU guide camera, and divide by the total aperture flux measured from the same FAU image. The distribution of these coupling estimates, shown in black in Figure \ref{fig:couple}, has a median of $\epsilon=0.46\%$, indicating that the fiber coupling efficiency is close to the theoretical expectation given the actual image quality.  

We can compare this to the purely theoretical simulations presented in \citet{Shaklan1988}. Assuming the mean seeing during commissioning of 1.7$^{\prime\prime}$ is entirely due to atmospheric turbulence (so, neglecting the wind-induced seeing component), we should expect the Fried parameter at 850~nm to be r$_0\sim0.10$~m and therefore $D/r_0\sim7.0$. Figure 5 of \citet{Shaklan1988} shows that the estimated coupling efficiency with the telescope focal ratio matched to the fiber NA, light with a wavelength close to the single-mode cutoff of the fiber, high-speed tip/tilt guiding, and $D/r_0\sim7.0$ peaks at approximately $14\%$. This value falls off to approximately $5\%$ without high-speed guiding. While the optical parameters of our test system are similar to this simulated case, we do not have high-speed tip/tilt guiding, and our on-sky coupling efficiency is approximately a factor of 10 worse than the \citet{Shaklan1988} simulation. This is likely due to the combined effects of wind shake and atmospheric turbulence, which impacts the actual PSF shape and motion in a way that is not captured in the \citet{Shaklan1988} simulations but is captured in our coupling efficiency estimates based on directly integrating the observed PSF. Analyses of stellar images taken with a different instrument at MINERVA using a high-speed CMOS camera capable of gathering images at tens of Hz indicate that substantial image motion is present at high frequencies. We expect that improving the bandwidth of the tip/tilt correction to 20 Hz or more should significantly improve our coupling efficiency.

\begin{figure}[h!]
	\centerline{\includegraphics[width=85mm,height=60mm]{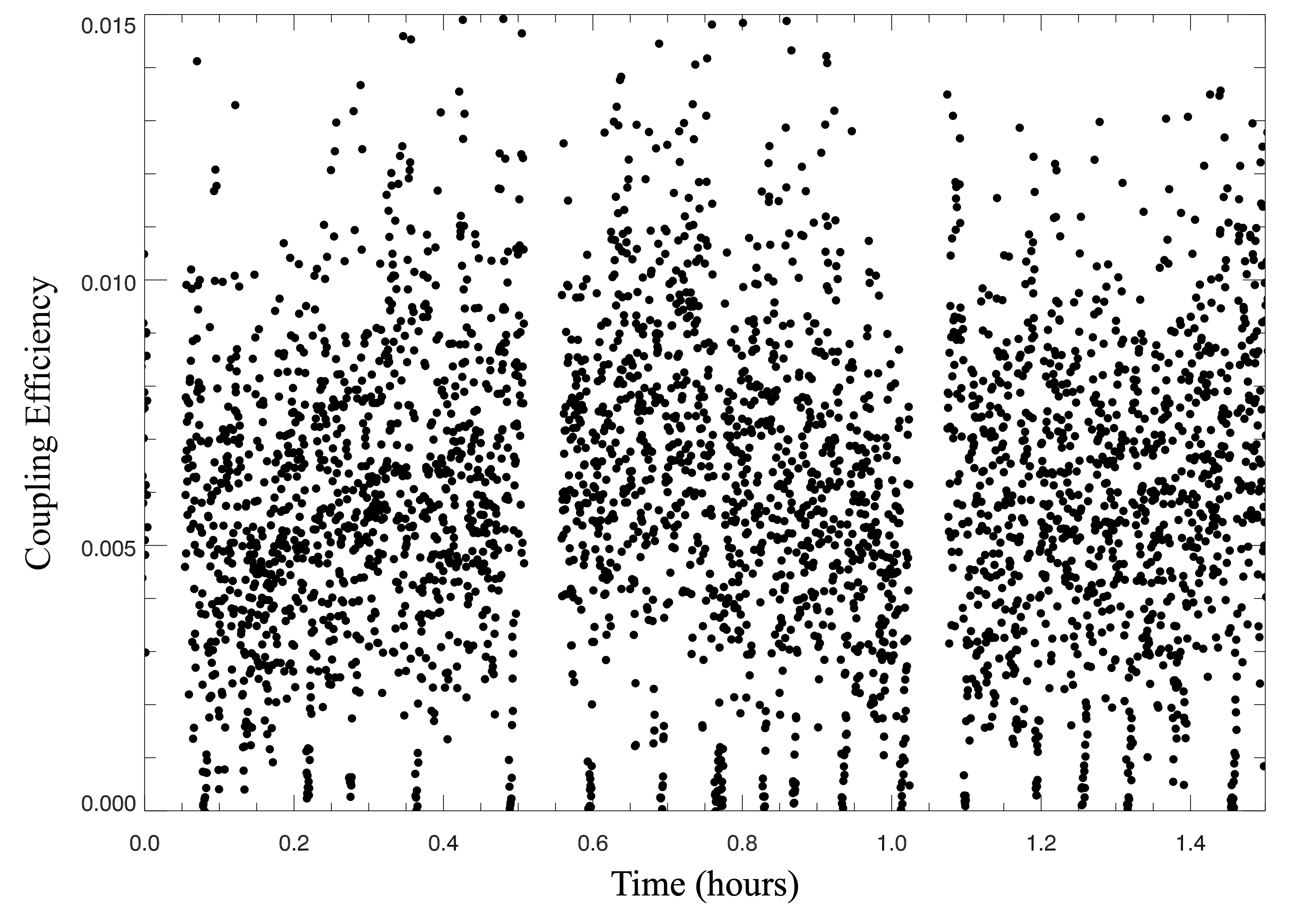}}
	\caption{Temporal variation in seeing-limited, single-mode fiber coupling efficiency observed with MINERVA-Red over a period of 1.5 hours. The coupling efficiency is highly variable, as is expected with seeing-limited coupling into single-mode fiber (see \citealt{Shaklan1988}). }
	\label{fig:time}
\end{figure}

\section{Discussion and Conclusions}

A high-resolution astronomical spectrometer designed to work with single-mode fiber can be small, inexpensive, and built from commercial, off-the-shelf parts. At the same time, such an instrument is immune to some of the important effects, such as modal noise, that may limit radial velocity precision with a multi-mode fiber instrument. Coupling starlight into such a small fiber is a challenge that requires an AO system on a large telescope. However, coupling light into the single-mode fiber is simplified if using a small telescope, where the ratio of the telescope diameter to the Fried parameter is small even at a site of moderate quality. We demonstrate coupling starlight into a single-mode fiber with a 0.7~m telescope under conditions of modest image quality with effective mean seeing of 1.7$^{\prime\prime}$. The MINERVA-Red FAU and spectrometer allow for simultaneous monitoring of a star using a guide camera and an exposure meter, which removes the sky conditions and telescope efficiency from the calculation of fiber coupling efficiency. After taking into account the reported efficiencies of the MINERVA-Red fiber system, optics, and detectors, we find that $0.45^{+0.3}_{-0.3}\%$ of starlight is coupled into the fiber using an image stabilization guiding system that operates at approximately 0.3 Hz and provides 0.5$^{\prime\prime}$ root-mean-square image centroid stability. As can be seen in Figure \ref{fig:time}, the coupling efficiency is highly variable on short timescales. This is expected with seeing-limited coupling into single-mode fibers since the coupling is dominated by speckles, which evolve on very short timescales. Variations of order unity are seen.

While this efficiency is low compared to the light-gathering capability of a larger, multi-mode fiber, it is sufficient to enable interesting science using small telescopes and inexpensive spectrometers. While we might think of the exoplanet discovery space for small telescopes as limited, there is rich science that can be carried out with sub-meter class telescopes. The spectral resolution element of an $R=10^{5}$ system at $\lambda=500$~nm is 0.005~nm, and the flux from a V=7 star in that one resolution element is $\sim800~$photon s$^{-1}$ m$^{-2}$. Assuming 0.5$\%$ coupling efficiency, a telescope with an effective collecting area of 0.5 m$^{2}$, and overall spectrograph efficiency of $10\%$, signal-to-noise of 50 can be achieved in a few hours of effective integration, which is useful for reconnaissance spectroscopy of exoplanet candidates and mass measurements for massive, short-period planets. 

Future work to improve the MINERVA-Red fiber coupling efficiency will focus on improving the image quality delivered by the FAU. A factor of two reduction in the image FWHM, which is consistent with other measures of site quality, should increase the average coupling efficiency, but image motion resulting from wind shake may limit those coupling improvements. However, the increased tip/tilt guiding bandwidth and low-order wavefront correction afforded by a low-cost adaptive optics system would significantly improve the delivered image quality and provide a significant boost to the overall MINERVA-Red efficiency. 

\section*{Acknowledgments}
We would like to thank an anonymous referee for comments and suggestions that helped to improve this manuscript. MINERVA is a collaboration between the Harvard-Smithsonian Center for Astrophysics, The Pennsylvania State University, the University of Montana, the University of Southern Queensland, and the University of Pennsylvania. MINERVA is made possible by generous contributions from its collaborating institutions and \fundingAgency{Mt. Cuba Astronomical Foundation}, \fundingAgency{ The David \& Lucile Packard Foundation}, the National Aeronautics and Space Administration \fundingAgency{NASA EPSCOR, ExEP, and Nancy Grace Roman programs} (NASA grants NNX13AM97A and NNX13AI79G), \fundingAgency{ The Australian Research Council} (ARC LIEF grant LE140100050), and the \fundingAgency{ National Science Foundation} (NSF grants 1516242 and 1608203). Funding for MINERVA data-analysis software development is provided through a sub-award under a NASA award NASA MT-13-EPSCoR-0011. This work was partially supported by funding from the Center for Exoplanets and Habitable Worlds, which is supported by the Pennsylvania State University, the Eberly College of Science, and the Pennsylvania Space Grant Consortium. 

\subsection*{Author Contributions}
Sliski played a lead role in assembling and testing the hardware described here and gathered the data used in these analyses. Blake analyzed the data. Eastman assisted with data analysis,  software for telescope operations, and hardware implementation. Eastman is the PI of the MINERVA project. Halverson  contributed expertise on the use of single-mode fibers.

\subsection{Financial Disclosure}

None to report.

\subsection{Conflict of Interest}

The authors declare no potential conflicts of interest.

\bibliography{main}%

\end{document}